\documentclass{elsart}

\usepackage{graphicx}
\usepackage{amsmath}

\def\av#1{\langle #1 \rangle}
\def\kc{k_{\rm c}}\def\l{\lambda}\def\rmax{r_{\rm max}}
\def\gsim{\mathrel{\raise.3ex\hbox{$>$\kern-.75em\lower1ex\hbox{$\sim$}}}}
\def\dmin{d_{\rm min}}

\begin{document}

\begin{frontmatter}

\title{ Geographical Embedding of Scale-Free Networks}

\author[adr1]{Daniel ben-Avraham \corauthref{cor1}},
\author[adr2]{Alejandro F. Rozenfeld},
\author[adr2]{Reuven Cohen},
\author[adr2]{Shlomo Havlin},
\corauth[cor1]{Corresponding author. E-mail: benavraham@clarkson.edu}
\address[adr1]{Department of Physics, Clarkson University, Potsdam, New~York
13699-5820, USA}
\address[adr2]{Minerva Center and Department of Physics, Bar-Ilan University, Ramat-Gan 
52900, Israel}

\begin{abstract}

A method for embedding graphs in Euclidean space is suggested.  The method
connects nodes to their geographically closest neighbors and economizes
on the total physical length of links.  The topological and geometrical
properties of scale-free networks embedded by the suggested algorithm are 
studied
both analytically and through simulations.  Our findings indicate dramatic 
changes
in the embedded networks, in comparison to their off-lattice counterparts, and
call into question  the applicability of off-lattice scale-free models to realistic,
everyday-life networks.

\end{abstract}

\begin{keyword}
Internet \sep scale-free \sep networks \sep embedding \sep lattice
\PACS 89.75.Hc \sep 05.50.+q \sep 89.75.Da 
\end{keyword}
\end{frontmatter}

\section*{\bf 1. Introduction}

The Internet and the World Wide Web (WWW), the electricity
power grid, networks of flight connections, of social
contacts, and neuronal networks of the brain are few of the
many examples of networks that surround us and that may be
usefully described as
graphs~\cite{bol,molloy,bar_review,dor_review,bornholdt}. 

Graph theory is rooted in
the 18th century, beginning with the work of Euler. 
Early efforts focused on properties of special (and usually
small) graphs. In the 1960s, Paul Erd\H{o}s
and Alfr\'ed R\'enyi~\cite{er,er2,er3} initiated the study
of {\em random graphs}, known also as ER graphs.
The unlimited size and randomness of ER graphs made them
natural contenders for models of large networks encountered in
everyday life.

In 1967 the psychologist Stanley Milgram asked himself how many
acquaintances, on average, connect between a person on the
East coast and another on the West coast of the USA. Following some 
research he finally concluded that the number of
acquaintances is surprisingly small, just about
six~\cite{milgram}.  This celebrated ``six degrees of
separation'' achieved widespread popularity.  It was later
realized that an average short path, typically of order $\ln N$ 
($N$ being the number of nodes, or the {\em size} of the network), connects
between randomly selected nodes in most naturally occurring networks.  
Recognizing
Milgram's contribution the effect came to be known as the {\em small world}
property of networks.  ER graphs possess the small world property.

However, an important ingredient was still missing.  In networks of social
contact the people that are connected to a certain
individual (such as family, acquaintances at work, and
friends) are highly likely to be connected among themselves. This high 
degree of {\em clustering} is absent in ER graphs. {\em Small World Networks}, 
introduced by Watts and
Strogatz~\cite{watts98,watts99}, exhibit high degrees of
clustering as well as the small world property. They mark an
important stage in the modeling of everyday networks and are
studied in their own right.  

It was recently realized that in 
addition to a high clustering
index and the small world property real life networks exhibit
yet a third important characteristic: a {\em scale-free}
degree distribution.  The {\em degree} of node $i$ is the
number of links, $k_i$, connected to the node.  The
likely degree of the various nodes in most observed networks
follows a power-law distribution:
\begin{equation}
\label{Pscale-free}
P(k)= ck^{-\lambda}\;,\qquad m\leq k\leq K\;,
\end{equation}
where $c$ is an appropriate normalization factor, $m$ is the
minimal degree of any given node, and
the cutoff degree
$K$ depends on the size of the network,
$K\sim m N^{1/(\lambda-1)}$~\cite{cohen,dor_cutoff}.
The term {\em scale-free} refers to the fact that the moments
$\av{k^n}$ for $n\geq\lfloor{\lambda}\rfloor$ do not exist(diverge), 
in the limit of $N\to\infty$.  Scale-free networks possess the small
world property, and it is possible to construct them with a high degree of
clustering.
Many important examples follow this pattern, including the
Internet~\cite{internet,fal} and the WWW (both for incoming
and outgoing links)~\cite{bar_degree,broder}, social networks,
and virtually any large network arising in some natural
context.  See~\cite{bar_review,dor_review}, for excellent
reviews, and~\cite{bornholdt}, for a timely anthology.  

An oft-neglected 
aspect in the modeling of everyday-life networks is the fact that
they are embedded in physical (Euclidean) space and possess a {\em geography},
in addition to their {\em topology}.  The spatial location of nodes and the 
length
of the connecting links is never a consideration in the models of 
networks discussed
above.  Nevertheless, routers of the Internet and social networks lay on the
two-dimensional surface of the globe; neuronal networks in brains occupy
three-dimensional space, etc.   The likelihood of connections between the 
nodes in
such networks is certainly affected by their geographical proximity, and one
expects nontrivial consequences arising from this interplay between 
geography and
topology~\cite{yook02}.  Yet, most studies so far have focused on topological
networks, where the nodes and links exist in some abstract space, devoid of
metric. 

In this paper we consider the embedding of networks in Euclidean
space.  In Section~\ref{general}, we discuss general aspects of embedding and
introduce a specific algorithm for the embedding of random networks of 
arbitrary
degree distribution. This algorithm favors connections to nearest nodes and
economizes on the total length of the links.  We show that infinite 
networks can be
thus embedded if and only if the degree ditribution has compact support, 
that is,
provided that there exists a sharp upper cutoff $K<\infty$ for the degree 
of any
given node.  If the distribution has no compact support, then 
embedding introduces
an artificial cutoff and is only possible in a restricted sense.  In
Section~\ref{scale-free} we apply the embedding algorithm to the widespread 
case of
scale-free networks and study key structural properties of the resulting 
lattices. We conclude with a discussion of open work and a comparison to 
an interesting
related algorithm, proposed by Warren, Sander, and Sokolov~\cite{warren},
in Section~\ref{discussion}. 
\section{The embedding algorithm}
\label{general}
\begin{figure}[ht] 
\vspace*{0.cm} 
\includegraphics*[width=0.5\textwidth]{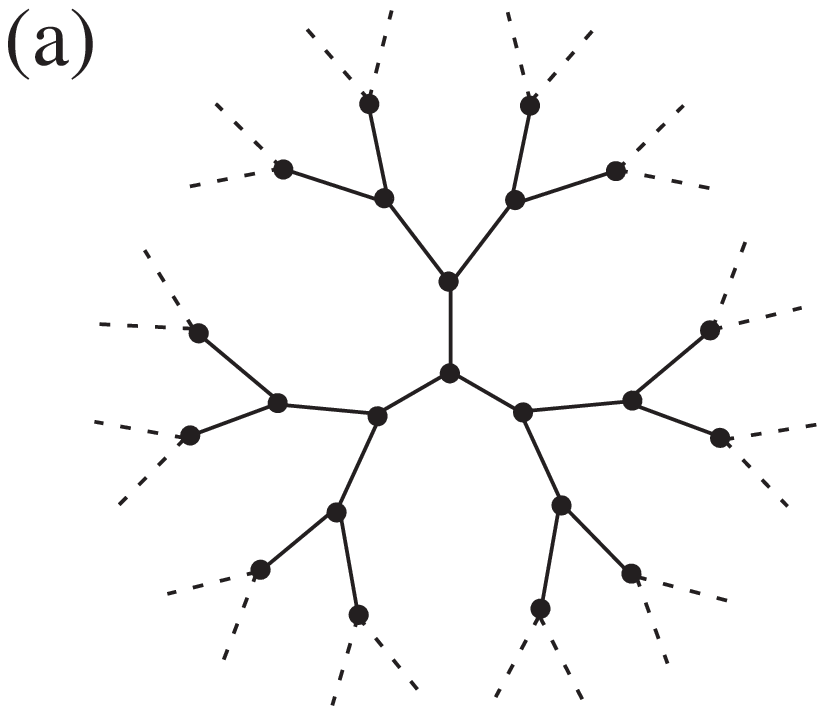} 
\vspace*{0.6cm} 
\includegraphics*[width=0.5\textwidth]{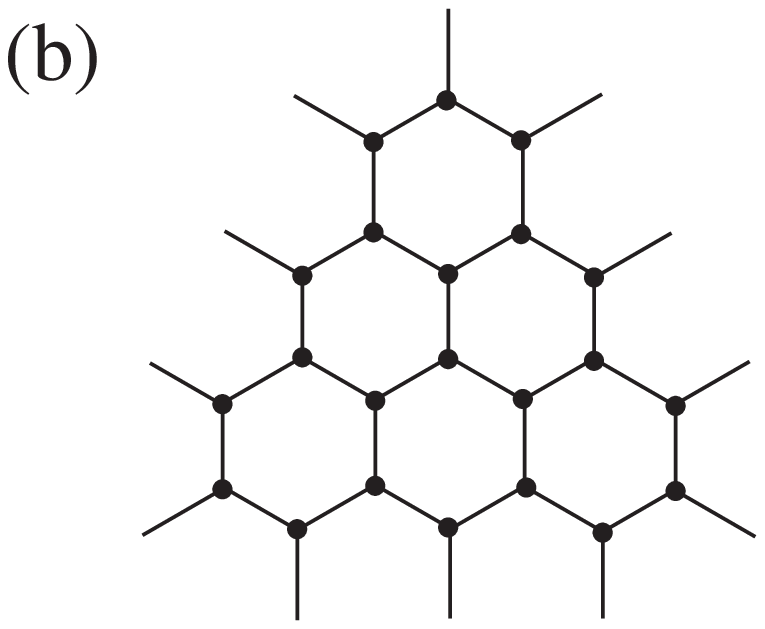}
\caption{Embedding problem.  The Cayley tree~(a) is a regular graph (all
nodes have the {\it same} connectivity, $k=3$) without loops.  It cannot be
embedded in Euclidean space, unless one allows for loops: 
The hexagonal lattice~(b)
is one example of embedding achieved in this way.  Notice that the degree
distribution remains unchanged.}
\label{cayley-hex}
\end{figure}

Our general problem is that of embedding a given infinite graph in Euclidean 
space. Suppose first that the graph contains no cycles (loops), i.e., it is 
a tree. Let
the volume of a node be $v$.  The total volume of nodes within chemical 
shell $l$
(nodes up to $l$ links away from a given node) increases as $\av{k}^{l}v$, 
where
$\av{k}=\int kP(k)\,dk$ is the average degree of the nodes. On the other 
hand, the
volume enclosed within a Euclidean distance $r$ scales as $r^d$ (in
$d$-dimensions).  Unless the links become progressively long, the exponential
growth in
$l$ cannot be sustained within the much slower algebraic
growth in $r$.  If the network is to be statistically
homogeneous, a lengthening of successive links is unacceptable. We 
conclude that infinite trees, such as the Cayley tree 
(Fig~\ref{cayley-hex}a),are not embeddable in Euclidean space~\cite{sbd}.  
Allowing for loops, the volume
constraint is reduced to the extent that embedding might be 
possible while retaining
homogeneity, and without a change in the degree distribution.  
An example is shown
in Fig~\ref{cayley-hex}b.

In order to make further headway we restrict ourselves to {\em random}
graphs, of arbitrary degree distribution $P(k)$, and the following embedding
algorithm~\cite{RCBH02}.  To each site $i$ of a $d$-dimensional lattice assign 
a
random connectivity $k_i$, taken from the degree distribution $P(k)$.  
Select site
$j$ at random and connect it to its closest neighbors, until its preassigned
connectivity
$k_j$ is realized,  or until all sites within distance
\begin{equation}
\label{r-k}
r(k_j)=Ak_j^{1/d}\;
\end{equation}
have been explored.  A link to site $l$ is allowed provided that: 
(a)~the site is
not saturated (its connectivity has not yet reached the preassigned $k_l$), 
and
(b)~its distance from site $j$ is smaller than $Ak_l^{1/d}$. 
Repeat for all sites. This algorithm makes sense in the context of 
social networks, where connections are
typically confined to one's immediate neighborhood.  The implied economy in the
physical length of links might render the algorithm useful for the 
modeling of other
networks of interest.

Suppose that one attempts to embed an infinite network, 
of degree distribution
$P(k)$, in an infinite lattice, by the above algorithm.  Nodes with a 
connectivity
larger than a certain cutoff $\kc(A)$ cannot be realized, because of the
possible saturation of surrounding sites.  Consider the 
number of links $n(r)$
entering a generic node from a surrounding neighborhood of radius $r$.  
Sites at distance $r'$ are linked to the node with probability 
\[Pr(k>(r'/A)^d)=\int_{(r'/A)^d}^{\infty}P(k')\,dk'\;.\]
Thus,
\[n(r)=\int_0^rdr'\,S_dr'{}^{d-1}\int_{(r'/A)^d}^{\infty}dk'\,P(k')\;,\]
where $S_d$ is the surface area of the $d$-dimensional unit sphere.
Reversing the order of integration and carrying out the spatial integral, we
obtain
\begin{eqnarray}
&&n(r)=\int_0^{(r/A)^d}dk\,P(k)\int_0^{Ak^{1/d}}dr'\,S_dr'{}^{d-1} 
\nonumber
\\
&&\ \ \ \ \ \ \ \ 
+\int_{(r/A)^d}^{\infty}dk\,P(k)\int_0^{r}dr'\,S_dr'{}^{d-1}    
\nonumber
\\&&=V_dr^d\Big\{\big(\frac{A}{r}\big)^d\int_0^{(r/A)^d}kP(k)\,dk  
+\int_{(r/A)^d}^{\infty}P(k)\,dk\Big\}\;,\nonumber
\end{eqnarray}
where $V_d=\frac{1}{d}S_d$ is the volume of the $d$-dimensional unit 
sphere. Finally, taking the limit $r\to\infty$, we derive the cutoff
\begin{equation}
\label{kc}
\kc=\lim_{r\to\infty}n(r)=V_dA^d\av{k}\;.
\end{equation}

We distinguish between two cases for the degree distribution $P(k)$: with and
without compact support.  The distribution $P(k)$ has compact support if there
exists a $K<\infty$ such that $P(k)=0$ for all $k>K$.  All finite networks 
have a
distribution with compact support, however infinite networks might too have 
compact
support (an example is the Cayley tree of Fig.~\ref{cayley-hex}a). 
If this is the
case, then $\av{k}\leq K$ is finite, and one can always select $A$ large 
enough so
that $\kc>K$.  In other words, {\em if there is compact support the 
network is
embeddable}.

If the network is infinite and without compact support, 
then if $\av{k}<\infty$, for any finite $A$ the cutoff $\kc$ is finite and 
the tail of the distribution for
$k>\kc$ is chopped off.  (Infinite networks with diverging $\av{k}$ are 
pathological--- we are not aware of any important practical application 
--- though they might
be embeddable.)  The cutoff connectivity $\kc$ implies a cutoff length
\begin{equation}
\xi=r(\kc)=(V_d\av{k})^{1/d}A^2\;.
\end{equation}  
The embedded network displays the original (chopped) distribution up to 
lengthscale
$\xi$ and repeats, statistically, at lengthscales larger than $\xi$.  
Indeed, $\xi$
is analogous to the correlation length in percolation theory, above 
criticality,
where the infinite cluster is {\em fractal} for $r<\xi$ and is homogeneous for
$r>\xi$~\cite{stauffer,havlin,book}.  We emphasize that the cutoff $\kc$ is 
of consequence even if
the distribution $P(k)$ is narrow, that is, even if
$\int_{\kc}^{\infty}P(k)\,dk\ll1$.  
Indeed, even in such a case the lengthscale
$\xi$ is finite (and controlled mainly by~$A$).

In summary, finite networks (all practical situations) are always 
embeddable by ourproposed algorithm.  Infinite networks are strictly 
embeddable only if their degree
distribution has compact support.  Otherwise, they are embeddable in a 
restricted
manner, with a cutoff $\kc$ imposed by the embedding, and statistical 
repetition
at lengthscales $r>\xi\sim A^2$. 
\section{Embedding of scale-free networks}
\label{scale-free}
We now apply the embedding algorithm to the wide\-spread case of 
scale-free
networks, Eq.~(\ref{Pscale-free}). Generally, we consider
embedding in $d$-dimensional lattices of size $R$, though in the numerical
simulations shown below we limit ourselves to two-dimensional square lattices
with periodic boundary conditions.   Because the lattice has a finite number of
sites,
$N\sim R^d$, the degree distribution~(\ref{Pscale-free}) has compact
support~\cite{cohen,dor_cutoff}:
\begin{equation}
\label{K}
K\sim mN^{1/(\l-1)}\sim R^{d/(\l-1)}\;.
\end{equation}
This, in conjunction with~(\ref{r-k}), implies a natural cutoff length
\begin{equation}
\rmax=r(K)\sim AR^{1/(\l-1)}\;.
\end{equation}

\begin{figure}
\centering
\includegraphics[width=0.6\textwidth, angle=-90]{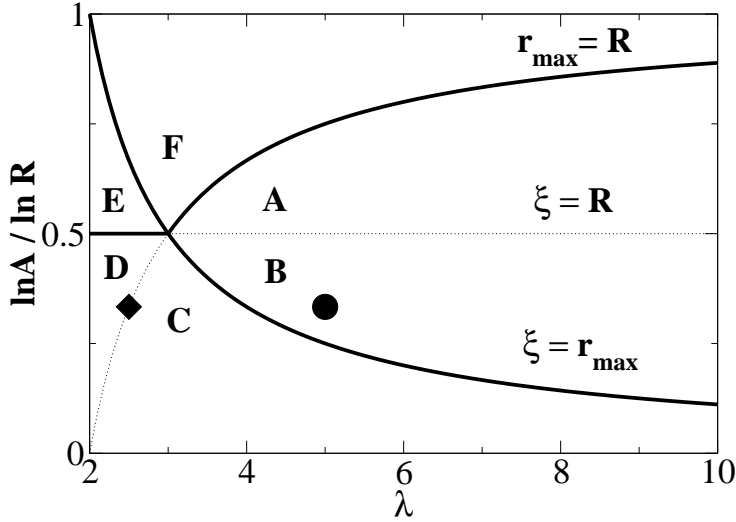} 
\caption{Regimes of embedded scale-free networks. 
A: $r_{max}<R<\xi$, B: $r_{max}<\xi<R$, C: $\xi<r_{max}<R$, 
D: $\xi<R<r_{max}$,E: $R<\xi<r_{max}$, F: $R<r_{max}<\xi$. 
The diagram can be reduced into just four regimes
separated by the cutoff $k_c$ and by finite-size effects. 
A and B: no cutoff
and no size effect; C and D: cutoff and no size effect; 
E: cutoff and size effect; F:
no cutoff but size effect. The two symbols indicate the parameters 
corresponding to Fig.~\ref{pinta}b, $\lambda=2.5$~(diamond)  
and $\lambda=5$~(circle). }
\label{regimes}
\end{figure}

The interplay between the three lengthscales, $R$, $\xi$, $\rmax$, 
determines the
nature of the network.  The embedding cutoff $\kc$ is imposed only if 
$\xi<\rmax$,
and there is statistical repetition for $r>\xi$.  Otherwise ($\rmax<\xi$), the
natural cutoff of $K$ is attained.  As long as
$\min(\rmax,\xi)\ll R$, the finite size of the lattice imposes no serious
restrictions.  Otherwise ($\min(\rmax,\xi)\gsim R$), finite-size effects become
important.   In short, there exist $6$ different regimes, characterized by 
the
relative ordering of $R$, $\xi$, and $\rmax$:
\smallskip
\noindent{Regime A:} 
$\rmax<R<\xi$. 
Natural cutoff $K$ is attained, and no finite-size
effects.  Statistical repetition occurs at lengthscales 
$r>\rmax$, rather than
$\xi$.
\smallskip
\noindent{Regime B:} 
$\rmax<\xi<R$. Same as regime A, but statistical
repetition occurs at lengthscales $r>\xi$.
\smallskip
\noindent{Regime C:} 
$\xi<\rmax<R$. Cutoff $\kc$ is imposed; nofinite-size effects.
\smallskip
\noindent{Regime D:} 
$\xi<R<\rmax$. Same as regime C.
\smallskip
\noindent{Regime E:} 
$R<\xi<\rmax$. Cutoff $\kc$ is imposed; strong
finite-size effects prevent statistical repetition.
\smallskip
\noindent{Regime F:} 
$R<\rmax<\xi$. Natural cutoff $K$ is attained;
strong finite-size effects.
\smallskip
\noindent
The various regimes are demarcated by the lines 
\begin{eqnarray}
\frac{\ln A}{\ln R}=\left\{ 
\begin{array}{ll}\frac{1}{2}+\frac{\l-3}{2(\l-1)}, &\rmax=R,
\\\frac{1}{2}, &\xi=R,
\\\frac{1}{2}-\frac{\l-3}{2(\l-1)}, 
&\xi=\rmax,\end{array} \right.
\end{eqnarray}
as shown in Fig.~\ref{regimes}.
\begin{figure}
\includegraphics*[width=\textwidth]{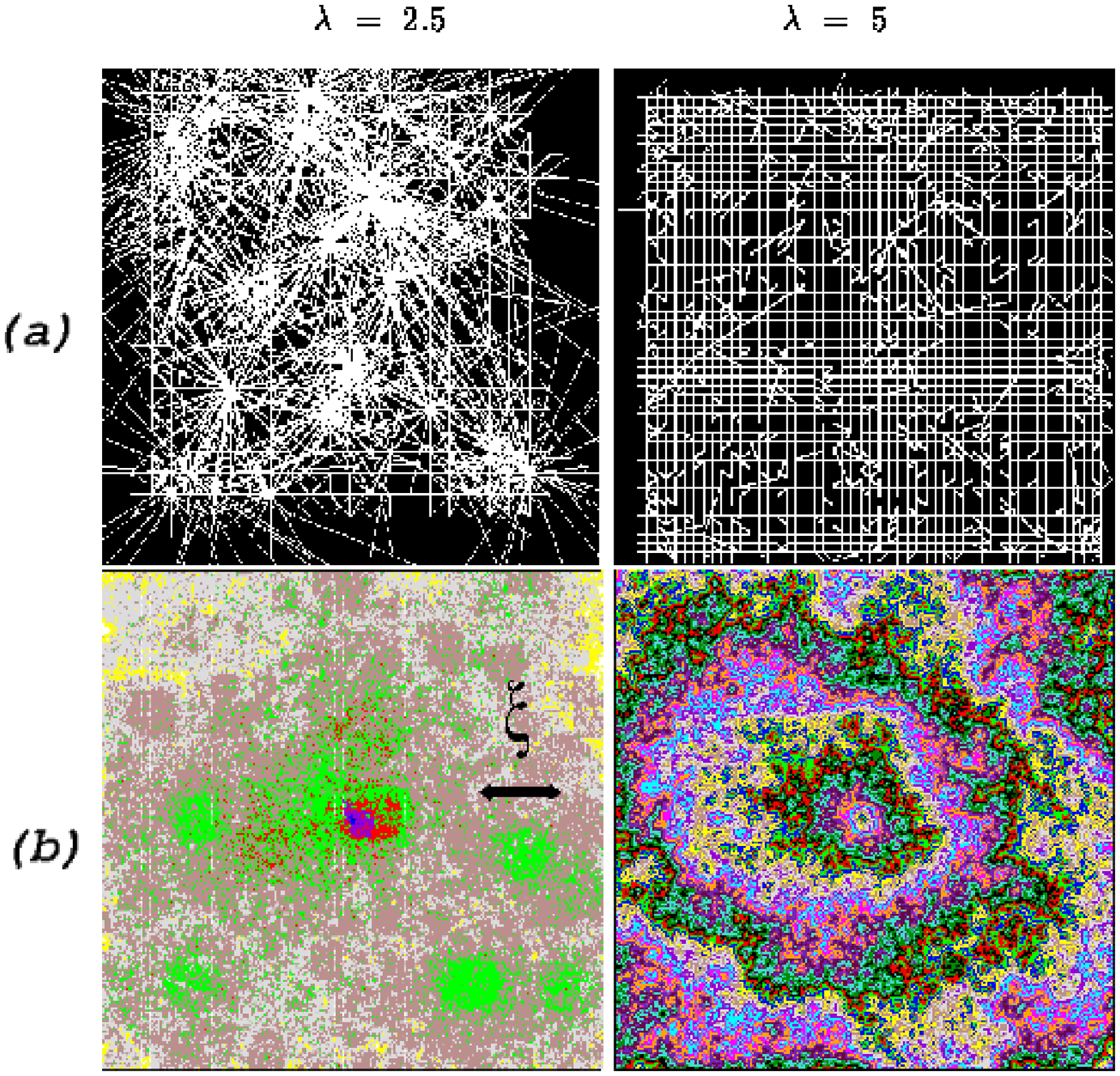}
\caption{Embedded scale-free networks. (a)~Actual networks (links 
are highlighted) for graphs with $\l=2.5$ and $5$,
embedded in $50\times 50$-sites square lattices.
(b)~Chemical shells, for the same parameters as in~(a), 
but for lattices of size
$300\times300$.  Notice the statistical repetition at lengthscales 
greater than$\xi$, for $\l=2.5$.}
\label{pinta}
\end{figure}
In Fig.~\ref{pinta}a, we present typical networks that result from our 
embedding
method, for $\l=2.5$ and $\l=5$.  Long range links become more noticeable 
as $\l$
decreases. In Fig.~\ref{pinta}b, we show the same networks as in
Fig.~\ref{pinta}a,
where successive chemical shells are shaded differently.  
Chemical shell
$l$ consists of all sites that can be reached by a minimal 
number of $l$ connecting
links relative to a given site (the central site, in the figure).  
For our choice
of parameters, $\l=5$ falls in the region of 
$\xi>\rmax$, while for $\l=2.5$,
$\xi<\rmax$.  In the latter case, we clearly see the statistical 
repetition beyond
the lengthscale $\xi$ (Fig.~\ref{pinta}b, $\l=2.5$). 
\begin{figure}
\includegraphics*[width=0.35\textwidth,angle=-90]{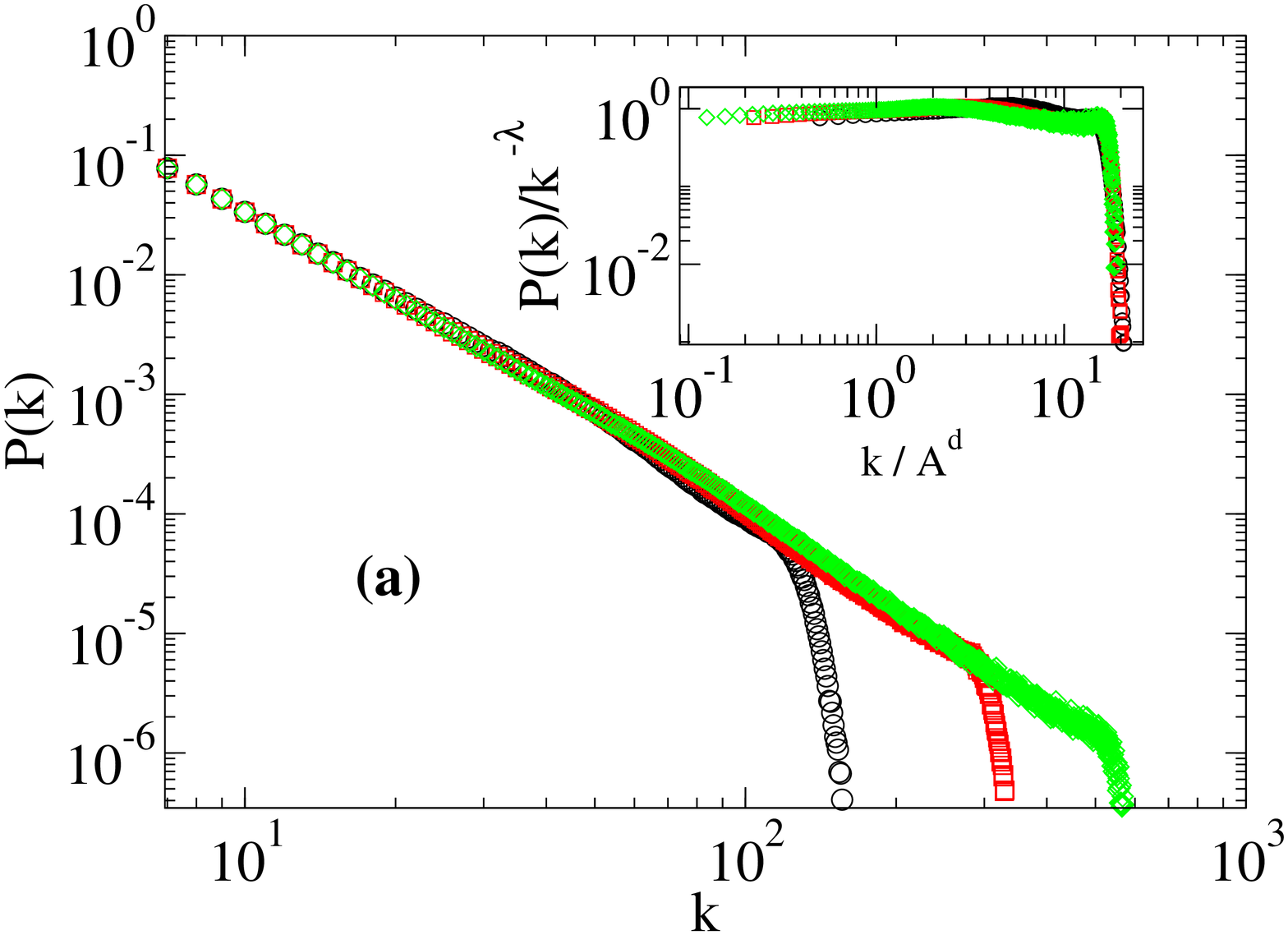}
\includegraphics*[width=0.35\textwidth,angle=-90]{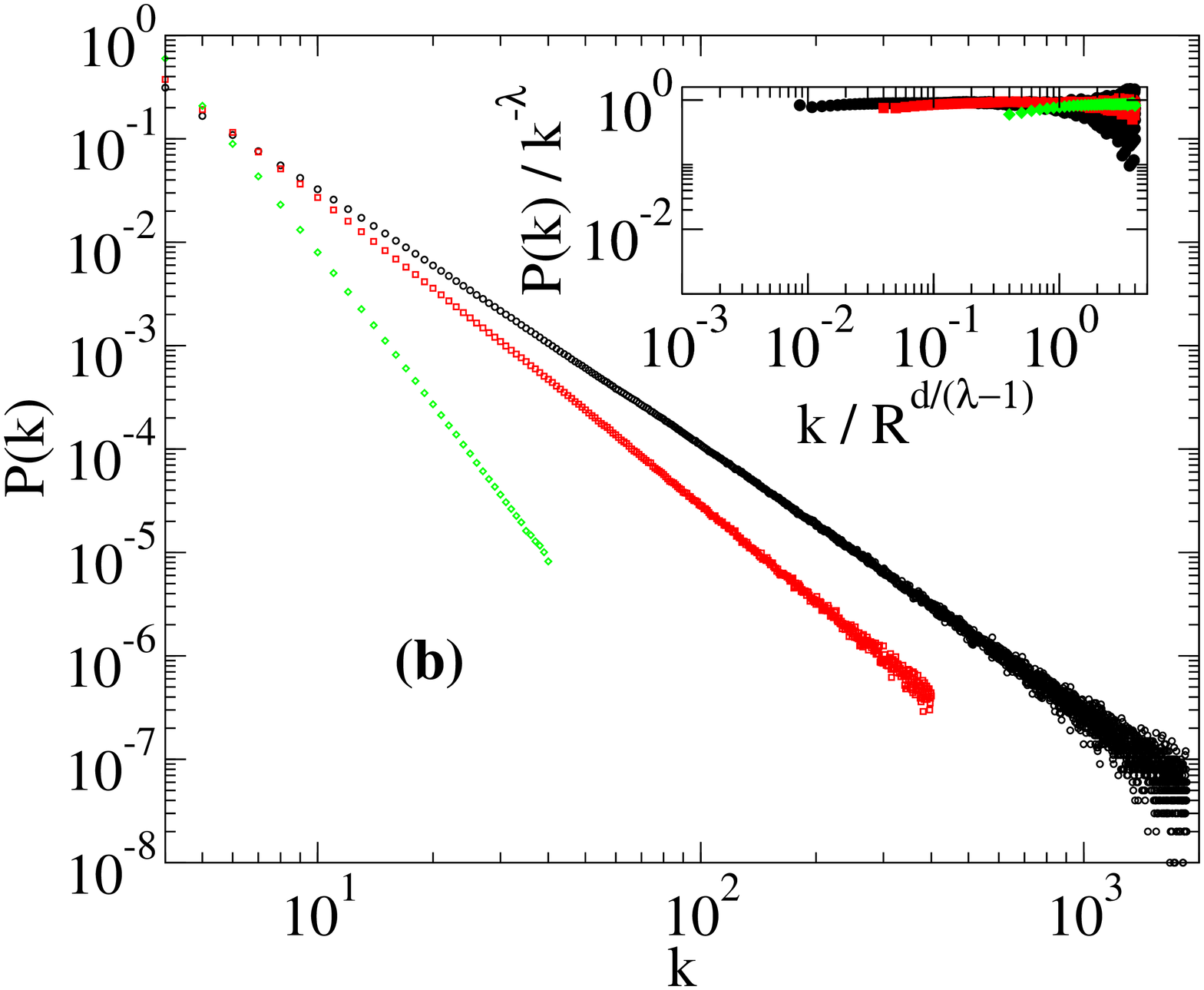}
\caption{Degree distribution of embedded scale-free networks,
(a)~when the cutoff
$\kc$ is imposed, and (b)~when the natural cutoff $K$ is achieved.  
For (a) the
lattice size is $R=400$, the distribution exponent 
is $\l=2.5$, and $A=2$
(circles), $3$ (squares), $4$ (diamonds).  The inset confirms the 
scaling of
Eq.~(\ref{kc}).  For (b) $R=100$, $A=10$, and $\l=2.5$ 
(circles), $3.0$ (squares),$5.0$ (diamonds). The inset confirms the 
scaling of Eq.~(\ref{K}).}
\label{distrib}
\end{figure}

The degree distribution of the embedded networks is illustrated in 
Fig.~\ref{distrib}.  In Fig.~\ref{distrib}a, various values of $A$ were 
selected
such that $\xi<\rmax$, and the distribution terminates at the imposed cutoff
$\kc$.  The scale-free distribution is altered slightly, for $k<\kc$, due to
saturation effects, but the overall trend is highly consistent 
with the original
power-law.  The data-collapse shown in the inset confirms the 
power-law, as well as
the scaling of the cutoff $\kc\sim A^d$.  In Fig.~\ref{distrib}b, $A=10$ was 
fixed
and values of $\l$ were selected such that $\xi>\rmax$.  The natural cutoff 
$K$ is
now attained, and the data collapse at the inset confirms the power-law
distribution as well as the known relation $K\sim mR^{d/(\l-1)}$.

We now turn to the relation between the
Euclidean metric and chemical length in our embedded scale-free networks.  The
chemical distance
$l$ between two sites is the minimal number of links $l$ connecting 
the two.  Thus,
if the Euclidean distance between the sites is $r$, then 
\begin{equation}
l\sim r^{\dmin}
\end{equation}
defines the
minimal (chemical) length exponent $\dmin$.  In 
order to compute $\dmin$ we regard the chemical shells as being roughly
smooth, at least in the regime $\xi>\rmax$, as suggested by 
Fig.~\ref{pinta}b($\l=5$).  Let the width of shell $l$ be 
$\Delta r(l)$, then
\begin{equation}
\label{lint}
l=\int dl=\int\frac{dr}{\Delta r(l)}\sim r^{\dmin}\;,
\end{equation}
since $\Delta l=1$.  The number of sites in shell $l$, $N(l)$, is, on 
the one hand,
$N(l)\sim r(l)^{d-1}\Delta r(l)$.  On the other hand, since the maximal
connectivity in shell $l$ is $K(l)\sim N(l)^{1/(\l-1)}$, the thickness 
of shell
$(l+1)$ is $\Delta r(l+1)\sim r(K(l))\sim AK(l)^{1/d}$.  
Assuming (for large $l$)
that $\Delta r(l+1)\sim \Delta r(l)$, we obtain
\begin{equation}
\Delta r(l)\sim r^{(d-1)/[d(\l-1)-1]}\;.
\end{equation}
Using this expression in~(\ref{lint}) yields
\begin{equation}
\label{dmin}
\dmin=\frac{\l-2}{\l-1-1/d}\;.
\end{equation}
Thus, for $d>1$, the dimension $\dmin<1$.  This result is opposite 
to all known
disordered media, where $\dmin\geq1$.  (A particularly simple example is
provided by a polymer chain, or its self-avoiding-walk model, 
where the end-to-end
distance is well-approximated by the Flory relation: 
$r\sim l^{3/(d+2)}$, i.e.,$\dmin=(d+2)/3$.)
We have also computed the fractal dimension of the 
newtorks and found that
the interior of the $l$-clusters is compact, with $d_f=d$~\cite{RCBH02}.
(Fig.~\ref{pinta} suggests that the hull of the clusters is fractal.)  
This resultwas confirmed by simulations.  It then follows 
that the fractal dimension of the
network in chemical space is also anomalous: 
$d_l=d_f/\dmin=d/\dmin>d$ (for $d>1$).
\begin{figure}
\includegraphics*[width=0.4\textwidth, angle=-90]{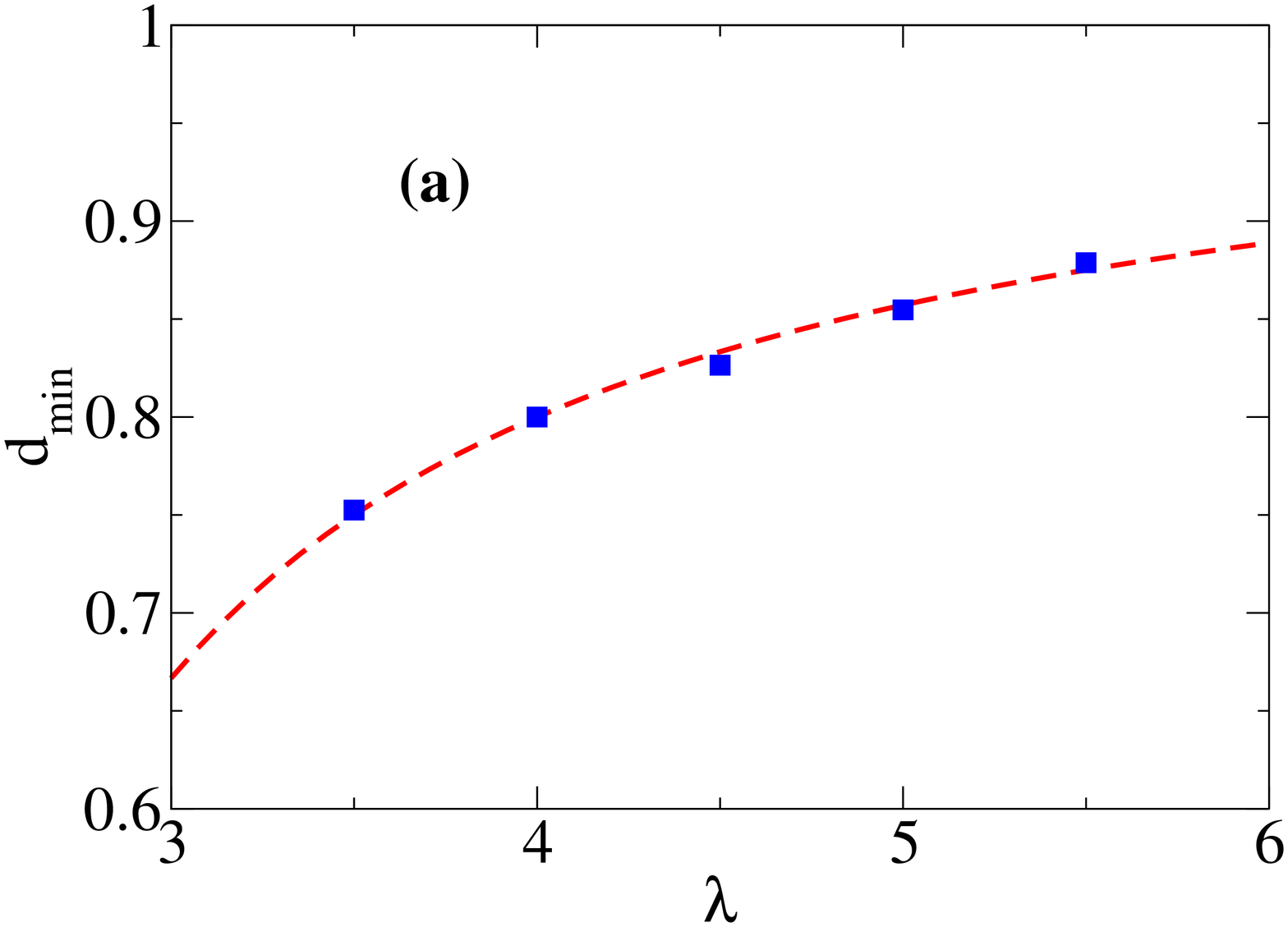} 
\includegraphics*[width=0.4\textwidth, angle=-90]{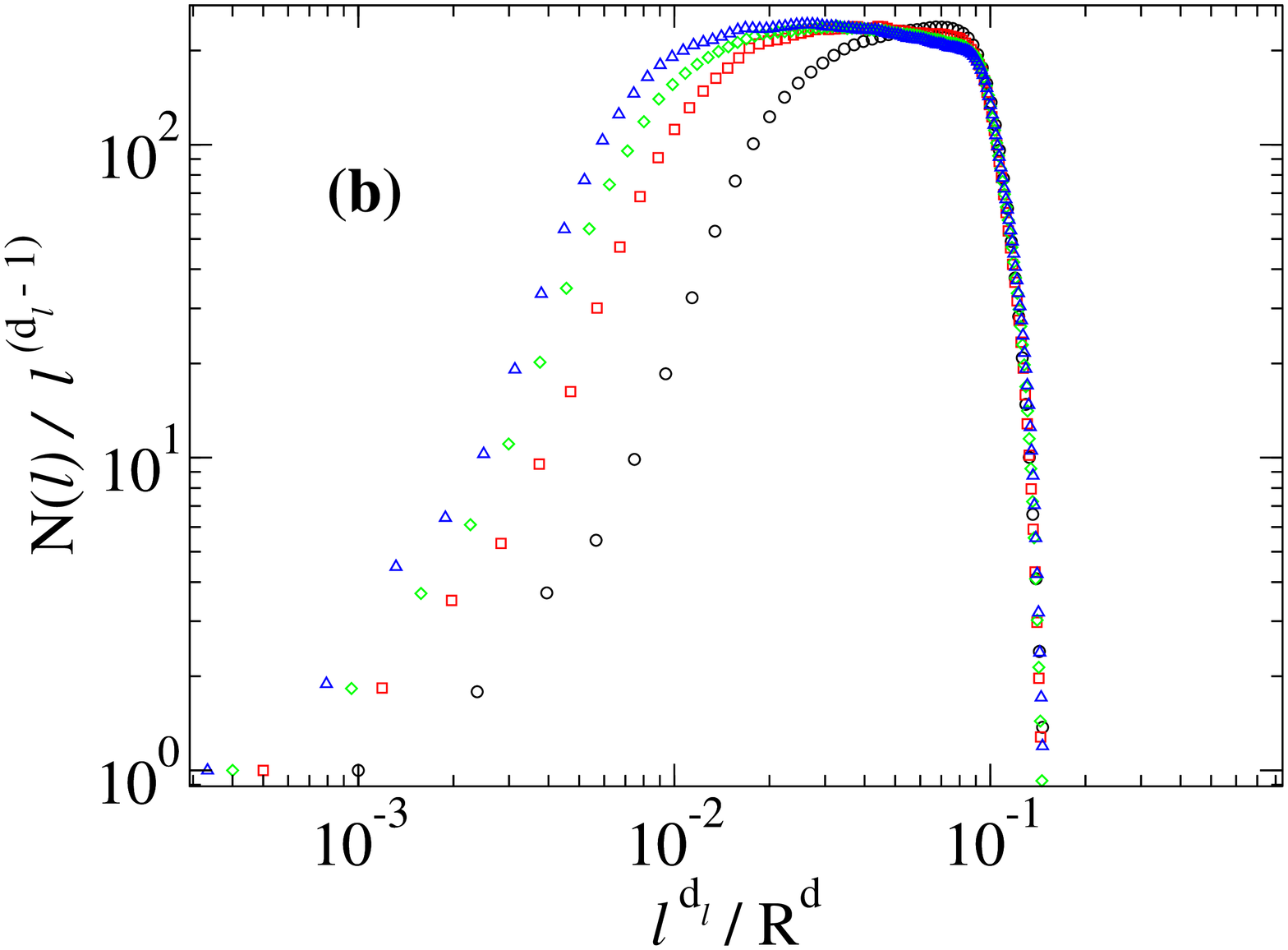} 
\caption{Minimal length exponent $\dmin$. 
(a)~$\dmin$ vs. $\l$. The analytic result
of Eq.~(\ref{dmin}) (curve) matches nicely the results 
measured from simulations(squares).  
(b)~Data colapse for the scaling function
$\Phi(l^{d_l}/R^d)$, for $\lambda=4$ and lattice sizes $R=1000$ 
(circles), $2000$ (squares), $2500$ (diamonds) and $3000$
(triangles). } 
\label{d_min}
\end{figure}
\noindent

In Fig.~\ref{d_min}a, we plot $\dmin$ as measured from simulations, and 
compared
with the analytical result of Eq.~(\ref{dmin}).  The scaling suggested in
Fig.~\ref{d_min}b, $N(l)\sim l^{d_l-1}\Phi(l^{d_l}/R^d)$, is valid only for
$\xi>\rmax$.  For $R\to\infty$, we expect that the network is 
scale-free up to
lengthscale $\xi$ and the analogous scaling would be 
$N(l)\sim l^{d_l-1}\Psi(l^{d_l}/\xi^d)$, where 
$\Psi(x\gg1)\sim x^{(d-d_l)/d_l}$.  It remains
an open challenge to conceive of a general scaling relation that would 
encompass all
of the regimes A -- F.
\section{Discussion}
\label{discussion}
In summary, we propose an algorithm for the embedding of networks in Euclidean
space.  Finite networks can always be embedded in this way.  
Infinite networks are
strictly embeddable only if the degree distribution has compact support.  
If not,
an artificial cutoff $\kc$ results from our embedding technique, 
and the networkrepeats itself, in a statistical sense, at lengthscales 
greater than $\xi(\kc)$.
We have applied our embedding algorithm to scale-free graphs, and studied the
resulting networks.

Concurrently with our work (and independently), Warren, Sander and
Sokolov~\cite{warren} proposed an embedding algorithm very similar to 
ours.  The
main difference in their approach is that connecting 
to saturated sites is allowed,
and a node is connected to as many of its  closest neighbors as necessary, 
until its
target connectivity is fulfilled.  In this case, our 
computation of $\kc$ in
Section~\ref{general} is still valid, only that now it is a 
{\em lower} cutoff for
the connectivity of the sites.  On the other hand, there
is no restriction, in this version of embedding, on the upper cutoff of the
distribution.  We anticipate that $\xi$ here plays the role of a 
`geometrical' lengthscale, similar to the lattice spacing for a 
critical percolation cluster grown
on a lattice. (The cluster is self-similar only at 
lengthscales larger than the
lattice spacing.)

Embedding scale-free networks in Euclidean space 
results in some dramatic changes
from the original graphs.  Consider, for example, the small 
world property, common
to off-lattice scale-free networks grown by any of the known techniques.  
The small
world property disappears at lengthscales greater than $\xi$: 
two sites separated
by a physical distance $r=n\xi$, $n\gg1$, would be connected by at least $n$
intervening links, due to the statistical repetition.  The number of 
sites withinradius $r$ is $N\sim n^d\xi^d$, so $l\sim N^{1/d}$.  Not 
surprisingly, this is the
same scaling as for Euclidean $d$-dimensional lattices.  
For distances $r<\xi$, wehave $l\sim r^{\dmin}\sim N^{\dmin/d}$, 
which represents small-world behavior, since
it improves on the Euclidean $l\sim N^{1/d}$, but a far cry from the
off-lattice $l\sim\ln N$.  Another
example of a dramatic change is the qualitative difference between 
percolation inoff-lattice and embedded networks.  Indeed, 
Warren et al.,~\cite{warren} find that
the percolation transition takes place in embedded lattices 
even for $\l<3$, in
contrast to off-lattice scale-free networks~\cite{bar_review,cohen}.
In view of
the fact that many everyday-life networks are 
embedded in physical space, we ought
to reconsider how the topological properties of 
scale-free graphs --- properties
that we normally attribute to the physical networks --- 
are affected by the
embedding.

\section*{Acknowledgements}
We are grateful to NSF grant PHY-0140094 (DbA) for partial support 
of this research.


\newpage
\begin{thebibliography}{00}

\bibitem{bol} B.~Bollob\'as, {\em Random Graphs}
   (Academic Press, London, 1985).
\bibitem{molloy}   M.~Molloy, B.~Reed:   {Random Structures and Algorithms} {\bf 6}, 161 (1995). 
\bibitem{bar_review}   R. Albert and A.-L. Barab\'asi,    {\it Rev. of Mod. Phys.} {\bf 74}, 47 (2002).
\bibitem{dor_review}   S. N. Dorogovtsev and J. F. F. Mendes,   {\it Adv. in Phys.}, {\bf 51} (4), (2002).
\bibitem{bornholdt}   S. Bornholdt and H.G. Schuster, eds.,   {\it Handbook of Graphs and Networks}, (Wiley-VCH, Berlin, 2003).
\bibitem{er}P. Erd\H{o}s and A. R\'enyi,{\it Publicationes Mathematicae} \textbf{6}, 290 (1959).
\bibitem{er2}P. Erd\H{o}s and A. and R\'enyi,{\it Publications of the Mathematical Institute of the Hungarian  Academy of Sciences} \textbf{5}, 17 (1960).
\bibitem{er3}P. Erd\H{o}s and A. R\'enyi{\it Acta Mathematica Scientia Hungary} \textbf{12},261 (1961).
\bibitem{milgram}S. Milgram, \textit{Psychology Today} \textbf{2}, 60 (1967).
\bibitem{watts99}D. J. Watts, \textit{Small Worlds}.(Princeton University Press, Princeton, 1999).
\bibitem{watts98}D. J. Watts and S. H. Strogatz,\textit{Nature} \textbf{393}, 440 (1998).
\bibitem{cohen}   R.~Cohen, K.~Erez, D.~ben-Avraham, and S.~Havlin,   {\it Phys.~Rev.~Lett.} {\bf 85}, 4626 (2000).
\bibitem{dor_cutoff}   S.~N.~Dorogovtsev, and J.~F.~F.~Mendes,   {\it Phys. Rev. E} {\bf 63}, 062101 (2001).
\bibitem{internet}   A. L. Barab\'asi and R. Albert,   {\it Science}, {\bf 286}, 509 (1999).
\bibitem{fal}   M.~Faloutsos, P.~Faloutsos, and C.~Faloutsos,   {\it Comput.~Commun. Rev.} {\bf 29}, 251 (1999).
\bibitem{bar_degree}   A.~-L.~Barab\'asi, R.~Albert, and H.~Jeong,   {\it Physica A}, {\bf 281}, 2115 (2000). 
\bibitem{broder}   A.~Broder, R.~Kumar, F.~Maghoul, P.~Raghavan, S.~Rajagopalan, R.~Stata,   A.~Tomkins and J.~Wiener,    {\it Comput. Netw.} {\bf 33}, 309 (2000).
\bibitem{yook02} S.-H. Yook, H. Jeong, A.-L. Barab\'asi, {\it Proc. Nat. Acad. Sci.} {\bf 99}, 13382 (2002).
\bibitem{warren}  C. P. Warren, L. M. Sander and I. M. Sokolov,   {\it Phys. Rev. E} \textbf{66}, 56105 (2002).
\bibitem{sbd} An amusing example is provided by starburst dendrimers (SBD) ---star-shaped polymers that grow in the patern of the Cayley tree ofFig.~\ref{cayley-hex}a.  The SBDs typically cannot grow beyond chemical shell$l=6$, due to lack of space.  See, for example, D.~ben-Avraham, L.~S.~Schulman, S.~H.~Bossmann, C.~Turro, and N.~J.~Turro,{\it J. Phys. Chem. B} {\bf 102}, 5088 (1998).
\bibitem{RCBH02} A. F. Rozenfed, R. Cohen, D. ben-Avraham and S. Havlin,{\it Phys. Rev. Lett.} \textbf{89}, 218701 (2002).
\bibitem{stauffer}   D.~Stauffer and A.~Aharony,   {\it Introduction to Percolation Theory}, 2nd edition    (Taylor and Francis, London, 1991). 
\bibitem{havlin}   A.~Bunde, and S.~Havlin (editors),   {\it Fractals~and~Disordered~System} (Springer, New~York, 1996).
\bibitem{book}   D. ben-Avraham and S. Havlin,    {\it Diffusion and Reactions in Fractals and Disordered Systems\/}   (Cambridge University Press, 2000).  

\end{thebibliography}
\end{document}